# A comparison between two OLS-based approaches to estimating urban multifractal parameters

Linshan Huang, Yanguang Chen

(Department of Geography, College of Urban and Environmental Sciences, Peking University, Beijing 100871, P.R. China. E-mail: chenyg@pku.edu.cn; linshanhuang@pku.edu.cn)

**Abstract:** Multifractal theory provides a new spatial analytical tool to describe urban form and growth, but many basic problems remain to be solved. Among various pending issues, the most significant one is how to obtain proper multifractal dimension spectrums. If an algorithm is improperly used, the parameter values will be abnormal. This paper is devoted to drawing a comparison between two OLS-based approaches for estimating urban multifractal parameters. Using observational data and empirical analysis, we will demonstrate how to utilize the double logarithmic linear regression to evaluate multifractal parameters. The OLS regression analysis has two different approaches. One is to fix the intercept to zero, and the other is not to fix it. The case studies show that the advisable method is to constrain the intercept to zero. The zero-intercept regression yields proper multifractal parameter spectrums within certain scale range of moment order, while the common regression results are not normal. In practice, the zero-intercept regression and the common regression can be combined to calculate multifractal parameters. Comparing the two sets of results, we can judge when and where multifractal urban structure appears. This research will inspire urban scientists to employ a proper technique to estimate multifractal dimensions for urban scientific studies.

# 1 Introduction

Urban fractal studies have emerged for more than 30 years, and many theoretical and empirical accomplishments were achieved. However, a great number of fundamental problems haven't been clarified yet. For instance, how to define the urban boundary for a study area? How to define the intercept of the regression model when dealing with the box-counting method? Many problems like those mentioned above are waiting to be solved. Fractal dimensions are the characteristic parameters of the models of fractal cities, which are used to replace the traditional measures such length, area, and density. A spatiotemporal comparison analysis of fractal dimensions will explain the evolvement of urban form and serves for theory construction and application in planning. Though fractal theory has primarily afforded modelling tools, the estimation of fractal dimensions requires concrete algorithms. One of the most common and most convenient algorithms is the ordinary least squares (OLS) regression analysis, which has two advantages. One is the integrity of data use because that the entire elements of diverse scales in a sample are taken into account (Chen, 2015), the other is the simplicity of operation since fractal dimensions can be easily calculated in common software such as Excel, SPSS and Matlab. Yet in estimating fractal dimensions with regression analysis, many particulars require particular attention. Those problems, if not properly solved, could cast a negative influence on the calculation and analysis results. Among them, the basic one is whether to fix the intercept to zero in the log-log linear regression modeling. In theory, the intercept of log-log linear function should be fixed to 0, corresponding to the proportional coefficient of power function equal to 1, but no empirical evidence of fractal studies is discovered and discussed in literature.

Previous studies on fractal cities and city fractals were mainly brought monofractal structure into focus. Monofractal analysis can merely yield a single fractal dimension value, and it is not sufficient to judge the advantages of an algorithm. Multifractals, however, is different. Cities and systems of cities in the real world can be treated as multifractals rather than monofractals. Multifractal method has been applied to the human geography, especially, urban studies. It can be used to characterize urban form (Ariza-Villaverde et al, 2013; Cavailhès et al, 2010; Chen and Wang, 2013; Frankhauser, 1998; Murcio et al, 2015), regional population (Appleby, 1996; Liu and Liu, 1993), urbanization (Chen, 2016), urban and rural settlement such as central places and rank-size distributions (Chen, 2014; Chen and Zhou, 2004; Haag, 1994; Hu *et al*, 2012 ; Liu and Chen, 2003). In general, two sets of fractal parameters are employed to describe multifractal cities, including global and local parameters. The global parameters include generalized correlation dimension and mass exponent, while the local parameters comprise singularity exponent and fractal dimension of fractal subsets in a complex fractal set (Chen, 2014; Feder, 1988). The basic global parameters are capacity dimension, information dimension, and correlation dimension (Chen and Wang, 2013; Murcio *et al*, 2015; Williams, 1997). In a given set of fractal dimensions, such as the set of generalized correlation dimensions, a clear numerical relationship exists between different dimension values. As a result, it is rather simple to conclude whether a fractal dimension value is reasonable or not. For example, if the information dimension value is greater than the capacity dimension value, the parameter relationship will be not correct. These characters can be employed to examine the effectiveness of multifractal spectrums.

This paper is devoted to research the proper method of evaluating multifractal parameters. The algorithm is based on the OLS calculation. By utilizing theoretical analysis and experimental results, we attempt to discuss whether to fix the intercept to zero when using OLS regression analysis to estimate multifractal parameters of urban form. The remaining parts are organized as follows. In Section 2, on the basis of theoretical analysis, it is recommended that a fixed intercept be used when computing the spectrum of multifractal parameters by means of OLS, otherwise a disorder might show up in the multifractal dimension spectra. In Section 3, based on the case study on Beijing's urban form, a comparison is made between the results of fixing-intercept treatment and floating-intercept treatment. In Section 4, several questions are discussed. Finally, this paper is concluded with a brief summarization of our study.

## 2 Theory and models

### 2.1 The condition of fractal dimension estimation

The definition of dimension is based on the power-law relationship between a measure and the corresponding scale (ruler length, yardstick). Generally, the scale-measure relationship can be expressed as below:

$$H(\varepsilon) = \varepsilon^{\pm d + h(\varepsilon)} = K(\varepsilon)\varepsilon^{\pm d}, \tag{1}$$

where $H(\varepsilon)$ represents some numerical quantity (measure) such as area $A(\varepsilon)$ and number $N(\varepsilon)$ which corresponds to certain linear size (scale) $\varepsilon$, $d$ refers to scaling exponent, and $h(\varepsilon)$ is a quantity related to proportionality coefficient $K(\varepsilon)$. Clearly, $K(\varepsilon)=\varepsilon^{h(\varepsilon)}$. Taking logarithms to a given base on both sides of equation (1) yields

$$\frac{\log H(\varepsilon)}{\log \varepsilon} = \pm d + h(\varepsilon), \quad (2)$$

which suggests that the precondition of $d$ representing a dimension is that $h(\varepsilon) \to 0$ when $\varepsilon \to 0$ (Fang, 1995; Mandelbrot, 1982). That is, the necessary condition of a valid dimension is

$$\lim_{\varepsilon \to 0} h(\varepsilon) = 0. \quad (3)$$

Thus, according to L'Hospital's rule (Appendix 1), we have

$$\lim_{\varepsilon \to 0} K(\varepsilon) = \lim_{\varepsilon \to 0} \varepsilon^{h(\varepsilon)} \to 1. \quad (4)$$

If so, equation (1) will change to a regular power law as follows

$$H(\varepsilon) = \varepsilon^{\pm d}, \quad (5)$$

which defines a valid dimension $d$. If $d$ is a fractional value rather than an integer, we will meet a fractal dimension (Mandelbrot, 1977; Batty and Longley, 1994). On the other hand, if the limit of $K(\varepsilon)$ is inexistent, equation (5) will be incompatible with the dimension estimation. In this case, the dimension calculation will result in what is called "lacunarity" (Mandelbrot, 1982). Empirically, the proportionality coefficient approaches to unit, that is $K(\varepsilon) \to 1$. This suggests, in practice, we can fix $K(\varepsilon)$ to 1 so that we will be able to have reasonable estimation of a fractal dimension value.

## 2.2 Proper multifractal parameter series

In order to characterize a multifractal pattern, two sets of parameters are required: One is the global parameters, and the other is the local parameters (Feder, 1988; Chen and Wang, 2013; Mandelbrot, 1999; Meakin, 1998; Stanley and Meakin, 1988). New alternative measure has been put forward (Grech, 2016). But the two sets of parameters are basic ones. Global parameters contain the generalized correlation dimension $D_q$ and the mass exponent $\tau_q$, and local parameters include the Lipschitz–Hölder exponent (singularity exponent) $\alpha(q)$, and the fractal dimension $f(\alpha)$ of the set supporting this exponent (Feder, 1988). The Legendre transform can be used to connect global parameters with local ones and transform one into the other (Badii and Politi, 1997; Feder, 1988).

The basic global parameter is the generalized correlation dimension $D_q$, which is based on the Renyi entropy and correlation function. This parameter can be expressed in the following form

$$D_q = -\lim_{\varepsilon \to 0} \frac{I_q(\varepsilon)}{\ln \varepsilon} = \begin{cases} \dfrac{1}{q-1} \lim_{\varepsilon \to 0} \dfrac{\ln \sum_{i=1}^{N(\varepsilon)} P_i(\varepsilon)^q}{\ln \varepsilon}, & (q \neq 1) \\ \lim_{\varepsilon \to 0} \dfrac{\sum_{i=1}^{N(\varepsilon)} P_i(\varepsilon) \ln P_i(\varepsilon)}{\ln \varepsilon}, & (q = 1) \end{cases}, \quad (5)$$

where $q$ is the order of moment ($-\infty < q < \infty$), $P_i(\varepsilon)$ denotes distribution or growth probability, indicating the ratio of the number of urban elements appearing in the $i$th box (rectangular grid) with linear scale $\varepsilon$ to the number of whole urban elements in the study area, $N(\varepsilon)$ refers to the number of nonempty box, $I_q(\varepsilon)$ represents the Renyi entropy. If $q=1$ as given, Renyi entropy will change into Shannon entropy. Another global parameter, the mass exponent $\tau_q$, can be given by

$$\tau(q) = D_q(q-1), \quad (6)$$

which suggests that the two global parameters can be converted into one another. As for the local parameters, the singularity exponent $\alpha(q)$ and the local fractal dimension $f(\alpha)$, can be derived by Legendre transform, that is

$$\alpha(q) = \frac{d\tau(q)}{dq} = D_q + (q-1)\frac{dD_q}{dq}, \quad (7)$$

$$f(\alpha) = q\alpha(q) - \tau(q) = q\alpha(q) - (q-1)D_q. \quad (8)$$

In theory, if we calculate the global parameters, we will evaluate the local parameters using Legendre transform, and *vice versa*.

It can be proved that the generalized correlation dimension is a monotonic decreasing function. Taking derivative of equation (5) with respect to $q$ yields

$$\frac{dD_q}{dq} = \begin{cases} -\frac{1}{(q-1)^2} \lim_{\varepsilon \to 0} \frac{\sum_{i=1}^{N(\varepsilon)} P_i(\varepsilon)^q \ln P_i(\varepsilon)}{\ln \varepsilon \sum_{i=1}^{N(\varepsilon)} P_i(\varepsilon)^q}, & (q \neq 1) \\ -\frac{d \sum_{i=1}^{N(\varepsilon)} P_i(\varepsilon) \ln P_i(\varepsilon)}{dq} = 0, & (q = 1) \end{cases} \qquad (9)$$

Apparently, $\ln P_i(\varepsilon) \leq 0$, $\ln(\varepsilon) \leq 0$, $\sum P_i(\varepsilon)^q > 0$, $(q-1)^2 \geq 0$, and thus $dD_q/dq \leq 0$. If $dD_q/dq=0$, then $D_q$ is constant, and the fractal is simple (monofractal); if $dD_q/dq<0$, then $D_q$ is a variable, and the fractal is complex (multifractals). Similarly, we can demonstrate that the mass exponent $\tau_q$ is a monotonic increasing function of $q$, the singularity exponent $\alpha(q)$ is a monotonic decreasing function of $q$, and the local fractal dimension $f(\alpha)$ is a first increasing and then decreasing function of $q$.

The range of the generalized correlation dimension depends on the embedding space. If a fractal city is examined in a digital map or remote sensing image, the Euclidean dimension of the embedding space is $d=2$. Thus the generalized correlation dimension of urban form based on the box-counting method should come between 0 and 2, i.e., $0<D_q<2$. If the moment order $q=0$, the local dimension, $f(\alpha)$, reaches its maximum value, the capacity dimension $D_0$. As a result, if the numerical quantity of $D_q$, $\alpha(q)$, or $f(\alpha)$ exceeds 2, it can be regarded as an abnormal value. Besides, as indicated above, $D_q$ and $\alpha(q)$ are monotonic decreasing functions, whereas $\tau_q$ is a monotonic increasing function of $q$ (Stanley and Meakin, 1988). The function $f(\alpha)$ corresponds to a unimodal curve, and the peak value is $\max(f(\alpha))=D_0$, which implies its monotonic increasing property when $q<0$ and its monotonic decreasing trend when $q>0$. Accordingly, those characteristics of global and local parameters can be employed to conclude whether or not the result is acceptable. Table 1 gives a plain view of the function property and value range of multifractal parameters for a regular multifractal object and sketches of their typical form.

**Table 1 Function property and value range of regular multifractal parameters**

| Parameter | Function property | Value range | Graph |
|---|---|---|---|
| $D_q$ | monotonic decreasing | 0~2 | Reverse S-shaped curve |
| $\tau_q$ | monotonic increasing | -∞~∞ | A curve comprising two straight lines |

| *α(q)* | monotonic decreasing | 0~2 | Reverse S-shaped curve |
| *f(α)* | monotonic increasing when *q*<0, monotonic decreasing when *q*>0 | *f(α)*≤*D*₀ | Unimodal curve |

## 2.3 Two approaches to evaluating fractal dimension

The general definition of dimension can be applied to fractal dimension. In other words, fractal parameters can be defined by a power-law relation between linear scales and the corresponding measures. For fractals, equation (1) should be replaced by

$$H(\varepsilon) = K\varepsilon^{-D}, \qquad (10)$$

in which $D$ denotes fractal dimension, $K$ represents proportionality coefficient. In theory, $K=1$. Taking logarithms on both sides of equation (10) yields a linear relation

$$\log H(\varepsilon) = \log K - D\log \varepsilon. \qquad (11)$$

Thus, the linear regression based on the least squares calculation can be employed to estimate the fractal dimension $D$. We have two approaches to making the regression analysis. One is to constrain $K$ to equal 1 by letting the constant term equal 0 ($\log K=0$), and the other is to let $K$ alone. For monofractal analysis, the use of the constraint condition, $\log K=0$, is not clear. However, for multifractal parameter estimation, the constraint condition is not dispensable. If $K=1$, we have

$$\log H(\varepsilon) = -D\log \varepsilon, \qquad (12)$$

which can be used to make zero-intercept regression analysis.

In practice, the calculation results of multifractal spectrums depend on the way of log-log linear regression. Empirical analyses of urban form show that a fixed intercept ($\log K=0$) will bring about a reasonable spectrum of multifractal dimensions when estimating fractal dimensions of urban form with OLS regression. In contrast, if the intercept is unfixed to zero ($\log K \neq 0$), the results of multifractal spectra may be disordered, or even unacceptable. The generalized dimension $D_q$ is a monotonically decreasing function that declines with the increase of order of moment $q$. Therefore, in a logical generalized dimension spectrum, the capacity dimension $D_0$ is bound to be greater than information dimension $D_1$, and information dimension $D_1$ is greater than correlation dimension $D_2$. In other words, the higher the moment order $q$ is, the smaller is the global dimension value. Otherwise, the spectrum might be abnormal, suggesting that something goes wrong with either the study object (such as internal disorder in a city system) or the estimation approach (such as an improper algorithm). Partial processes and results of the empirical analyses will be reported in the next section.

## 3 Empirical evidences

### 3.1 Study area and methods

In order to make a comparison between the effect of zero-intercept regression and that of the common linear regression, we apply multifractal measurements to the multiscaling description of the urban form of Beijing, the capital city of China. The total area of Beijing municipality is about 16800 km$^2$ and the metropolitan area is almost 13800 km$^2$. According to the sixth population census, the population of Beijing metropolitan area is approximately 18.8 million in 2010. Remote sensing images from 13 different years are available for analyzing the evolution of Beijing urban form, namely 1984, 1988, 1989, 1991, 1992, 1994, 1995, 1996, 1998, 1999, 2001, 2006, and 2009. The original ground resolution of those images is slightly different. For the sake of spatial comparability, the resolution of all the urban images is adjusted to 33.3 meters before abstracting data.

The box-counting method is employed to compute multifractal spectrums in different years. The first step is to determine urban boundary. The boundary of an urban agglomeration can be identified by the "City Clustering Algorithm" (CCA) (Rozenfeld *et al*, 2008). The CCA is originally developed to define urban area beyond the scope of its administrative boundary based on spatial distributions of the population at a fine geographic scale. In this paper, the population maps are replaced by the maps of land use for urban construction, which are extracted from 13 remote sensing images by supervised classification and relevant manipulation. The main process of parameter estimation is as follows. Firstly, we arbitrarily choose an urban land cell. The random selection of the initial cell doesn't impact the result. Then we repeatedly annex nearest neighbors of the boundary cells, which also represent urban construction, to the boundary cells until all neighbors of the boundary stand for non-urban construction. Thus a cluster will emerge. Repeat this process until all urban land-use cells have been assigned to a cluster, which denotes the urbanized area. In the whole process, a proper neighborhood should be defined first. In this paper, we select the radius of 700 meters as a threshold to define the neighborhood. That is to say, as long as a cell is less than 700 meters from the boundary cell, it can be called its neighbor, even though in the 33.3-meter-cell grid, they are not visually adjacent to each other. If an urban boundary is outlined on a digital map, we will have an *urban envelope* for a city in a given year (Batty and Longley, 1994; Longley *et al*, 1991).

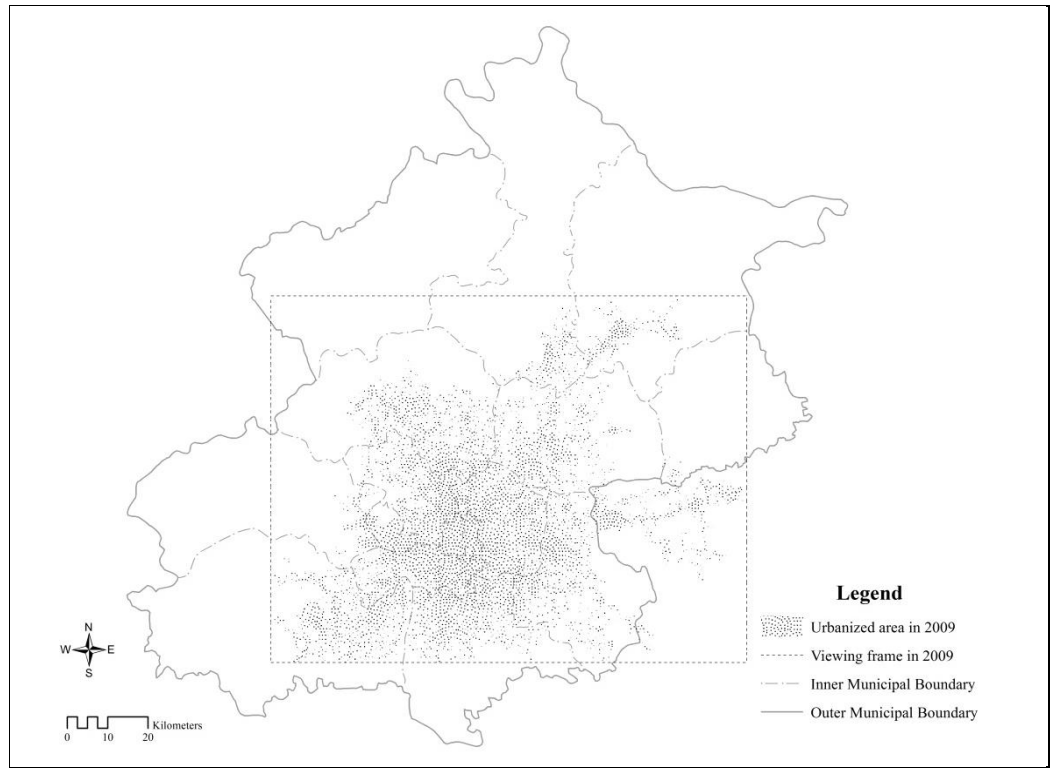

**Figure 1 Metropolitan area, urban agglomeration, and the viewing frame of Beijing in 2009**

To calculate fractal dimension of urban form using the box-counting method, a proper study area should be determined. For fractal dimension estimation based on digital maps, a study area can be compared to a viewing frame of photograph (Chen and Wang, 2013; Feng and Chen, 2010). There are two ways of defining study area of box-counting dimension (Chen, 2012). One is fixed study area (Batty and Longley, 1994; Shen, 2002), and the other is variable study area (Benguigui *et al*, 2000; Feng and Chen, 2010). Considering that the main aim of this article is not to compare the fractal dimension values from 1984 to 2009, we adopt a variable frame indicative of a variable study area for each year. It is easy to determine the viewing frame. Make a rectangular box as small as possible so that it just covers a given urban envelope. Thus we have a measure area of the urban envelope. Figure 1 illustrates the method and result of drawing the outline of study area from the metropolitan area of Beijing in 2009. Figure 2 shows partial urban agglomerations and corresponding viewing frames in the other 12 years. However, it is worth noticing that the sizes of urban agglomerations in earlier years derived from CCA aren't always smaller than those in later years. The error happens mainly in the process of extracting urban construction sites from remote sensing images. Many factors will lead to a deviation of data processing. For example, the quality of remote sensing images (image resolution, spatial resolution, shrouded by the cloud or not, covered with snow or not) is uneven, which cause inaccuracy in some extractions of urban land. Despite all that, the general trend of urban expansion can be clearly seen (Figure 2).

The extraction and processing of spatial data can be performed by means of ArcGIS technology and digital maps. By iteratively dividing the rectangle into four equal rectangles, side length $\varepsilon_k$ and the number of the nonempty rectangles $N(\varepsilon_k)$ in the $k$th step can be computed or counted to make datasets for each year. The datasets can be used to estimate multifractal parameters by implementing OLS calculation. The first step to make multifractal analysis of urban form is to distinguish the fractal structure types, that is to say, to judge whether the urban figure is a multifractal pattern. The basic criterion is the numerical relationship between the capacity dimension ($D_0$), the information dimension ($D_1$) and the correlation dimension ($D_2$). If $D_0=D_1=D_2$, the urban form is of monofractal (unifractal) structure; if $D_0>D_1>D_2$ significantly, the urban form takes on multifractal property. As indicated and demonstrated above, if $D_0<D_1$ or $D_1<D_2$ or $D_0<D_2$, there is something wrong (Chen and Wang, 2013).

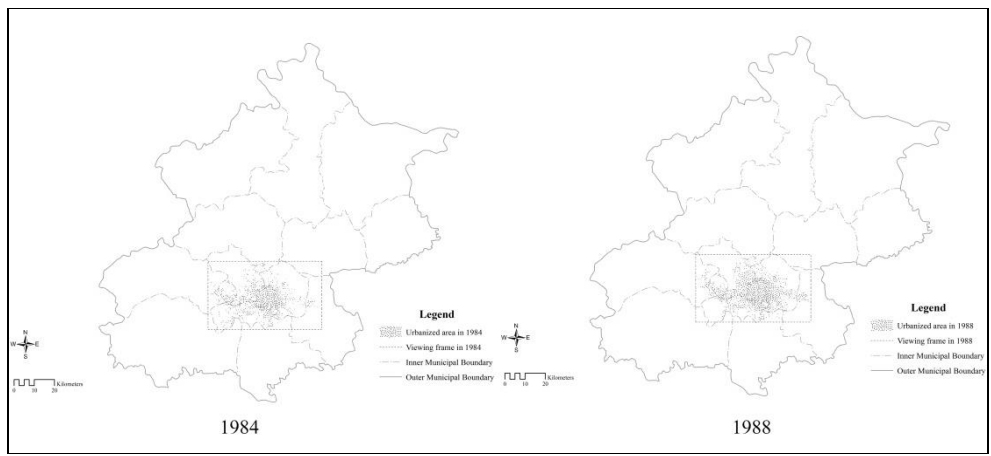

1984    1988

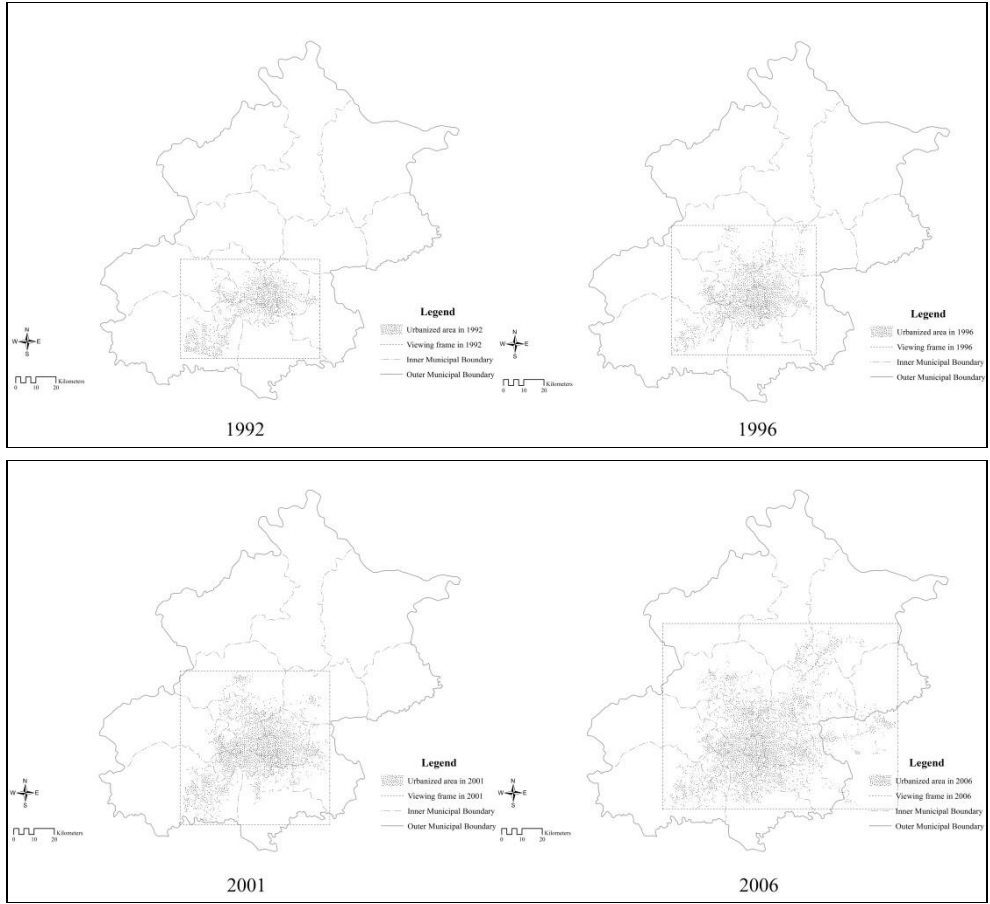

**Figure 2 Metropolitan areas, urban agglomerations, and the viewing frame of Beijing city from 1995 to 2006**

The process of calculation is as follows. First of all, the OLS regression is used to compute the three special fractal parameters by fixing the intercept to zero. It is easy to estimate $D_0$, $D_1$, and $D_2$ using box-counting method. The formula of capacity dimension is as below

$$D_0 = -\lim_{\varepsilon \to 0} \frac{\ln N(\varepsilon)}{\ln \varepsilon}, \qquad (13)$$

where $N(\varepsilon)$ denotes the number of non-empty boxes of size $\varepsilon$. The corresponding linear regression equation is

$$\ln N(\varepsilon) = \ln K - D_0 \ln \varepsilon, \qquad (14)$$

where $K=1$ and thus the intercept $\ln K=0$. The formula of information dimension is as below

$$D_1 = -\lim_{\varepsilon \to 0} \frac{I(\varepsilon)}{\ln \varepsilon} = -\lim_{\varepsilon \to 0} \frac{-\sum_{i=1}^{N(\varepsilon)} P_i(\varepsilon) \ln P_i(\varepsilon)}{\ln \varepsilon}, \qquad (15)$$

where $P_i(\varepsilon)$ denotes the "possibility" measurement of $i$th box of size $\varepsilon$, and $I(\varepsilon)$ refers to Shannon's information quantity. In the numerical sense, information quantity $I(\varepsilon)$ equals information entropy $H(\varepsilon)$, which is actually a type of spatial entropy in geographical analysis (Batty, 1974; Batty, 1976). The corresponding regression equation is

$$I(\varepsilon) = -\sum_{i=1}^{N(\varepsilon)} P_i(\varepsilon) \ln P_i(\varepsilon) = I_0 - D_1 \ln \varepsilon, \tag{16}$$

where the intercept $I_0=0$ in theory. The formula of correlation dimension is as below

$$D_2 = \lim_{\varepsilon \to 0} \frac{\ln \sum_{i=1}^{N(\varepsilon)} P_i(\varepsilon)^2}{\ln \varepsilon}, \tag{17}$$

The corresponding linear regression equation is

$$C(\varepsilon) = -\ln \sum_{i=1}^{N(\varepsilon)} P_i(\varepsilon)^2 = C_0 - D_2 \ln \varepsilon, \tag{18}$$

where $C(\varepsilon)$ denotes quadratic correlation function, and theoretically, the intercept $C_0=0$.

The next step is to estimate of the multifractal dimension spectra and the related parameters, including the global parameters $D_q$ and $\tau_q$, and the local parameters $\alpha(q)$ and $f(\alpha)$. After calculating $D_q$, we can compute $\tau_q$ using equation (6). Then, $\alpha(q)$ and $f(\alpha)$ can be worked out using Legendre transform, equation (7) and (8) (Appendix 2). In fact, it will be more convenient if we first compute the local parameters and then evaluate the global parameters using Legendre transform. The local parameters, $\alpha(q)$ and $f(\alpha)$, can be directly computed by $\mu$-weight method (Chhabra and Jensen,1989), and the formulae are as follows

$$\alpha(q) = \lim_{\varepsilon \to 0} \frac{1}{\ln \varepsilon} \sum_{i=1}^{N(\varepsilon)} \mu_i(\varepsilon) \ln P_i(\varepsilon), \tag{19}$$

$$f(\alpha) = \lim_{\varepsilon \to 0} \frac{1}{\ln \varepsilon} \sum_{i=1}^{N(\varepsilon)} \mu_i(\varepsilon) \ln \mu_i(\varepsilon), \tag{20}$$

where the $\mu$-weight is defined by

$$\mu_i(\varepsilon) = \frac{P_i(\varepsilon)^q}{\sum_i P_i(\varepsilon)^q}, \tag{21}$$

in which the probability measure is given by

$$P_i(\varepsilon) = \frac{M_i(\varepsilon)}{\sum_i M_i(\varepsilon)}, \tag{22}$$

where $M_i(\varepsilon)$ represents the area or pixel number of artificial construction appearing in the $i$th nonempty box. The regression equations are as follows

$$\sum_{i=1}^{N(\varepsilon)} \mu_i(\varepsilon) \ln P_i(\varepsilon) = \alpha(q) \ln \varepsilon, \qquad (23)$$

$$\sum_{i=1}^{N(\varepsilon)} \mu_i(\varepsilon) \ln \mu_i(\varepsilon) = f(\alpha) \ln \varepsilon, \qquad (24)$$

As indicated above, the constant term should be equal to zero; therefore, the intercept must be set to 0. Using equation (23) and (24), we can estimate the local parameters of multifractals using linear regression analysis (Figure 3). Then, using Legendre transform, we can estimate the global parameters and $D_q$ and $\tau_q$.

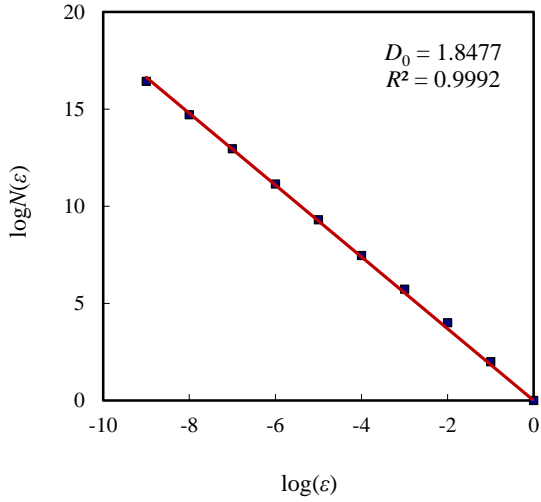

a. Capacity dimension ($q=0$, $D_0$)

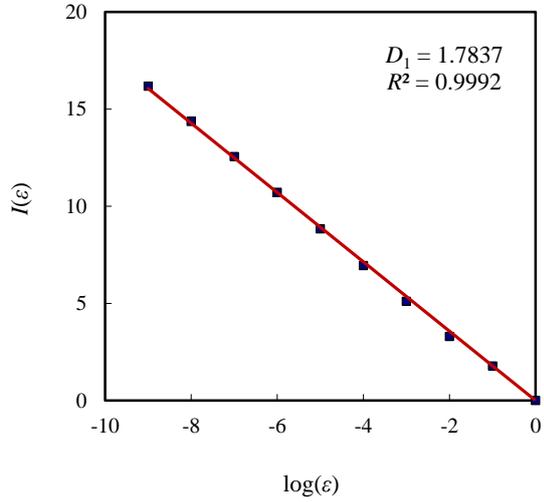

b. Information dimension ($q=1$, $D_1$)

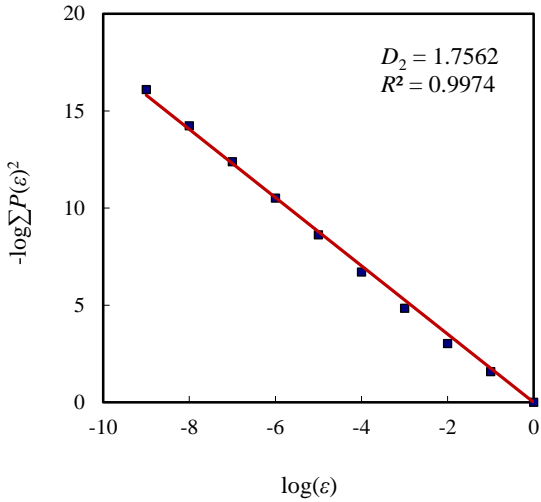

c. Correlation dimension ($q=2$, $D_2$)

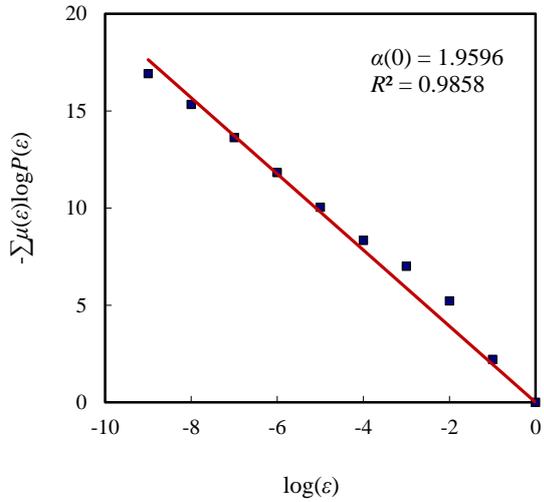

d. Singularity exponent ($q=0$, $\alpha(0)$)

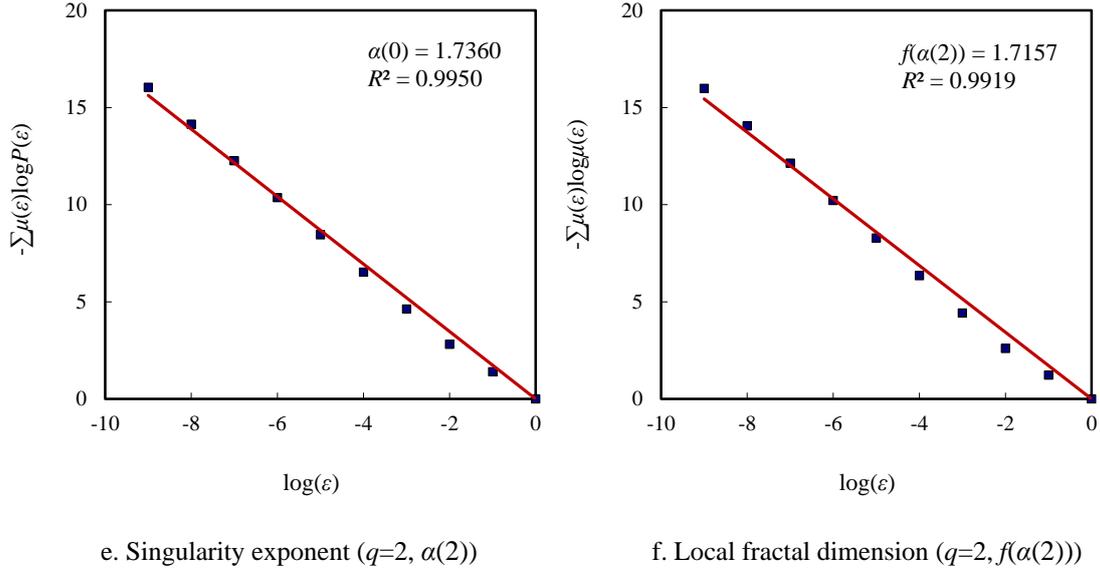

e. Singularity exponent ($q=2$, $\alpha(2)$)  f. Local fractal dimension ($q=2$, $f(\alpha(2))$)

**Figure 3 The log-log plots for evaluating multifractal parameters of Beijing's urban form in 2009 (examples)**

## 3.2 Calculations based on constrained intercept ($K=1$)

Based on the mathematical models shown above, the zero-intercept regression analysis can be made to estimate the capacity dimension, information dimension, and correlation dimension. Constraining the intercepts of regression equations with $\ln K=0$, $I_0=0$, and $C_0=0$, we can implement least square calculations using equations (14), (16), and (18). Table 2 displays the results of $D_0$, $D_1$, and $D_2$ from 1984 to 2009. It is evident that, through all the years, the inequality $D_0>D_1>D_2$ can always be established. Moreover, the relevant goodness of fit ($R^2$) is rather high, and all the values are higher than 0.995, i.e., $R^2>0.995$. A conclusion can be drawn that the urban form of Beijing bears multifractal pattern, and the urban growth is a multi-scaling process.

The results suggest that the multifractal properties of Beijing's urban morphology take on within certain range of scales. The parameter spectrums of 2009 year are displayed in Figure 4. The spectral curves of other years are similar to those in 2009. The generalized correlation dimension $D_q$ decreases monotonically. In the range of $q$ values from -0.5 to 20, the $D_q$ values are normal. However, when $q<-1$, the $D_q$ values exceeds the Euclidean dimension of the embedding space $d$. That is to say, $D_q>d=2$ for $q<-1$, and this is abnormal (Figure 4a). The graph of the mass exponent $\tau_q$ values manifests the monotonic increasing property. However, if $q<0$, an apparent bend can be examined (Figure 4b). The spectrum of $\alpha(q)$ reveals a similar shape and tendency to $D_q$ except that the value range is wider than $D_q$, but the $\alpha(q)$ curve becomes more abruptly than $D_q$ curve. This indicates that the $\alpha(q)$ value is more sensitive than the $D_q$ value. The curve is proper when $q>0$, but there is something wrong when $q<0$, Clearly, if $q<-1$, the $\alpha(q)$ value fails to converge (Figure 4c). The spectrum curves of the $f(\alpha)$ are normal when $q>0$, but it is not normal when $q<0$ because the local dimension values fail to approach to limits (Figure 4d). The relationship between $\alpha(q)$ and $f(\alpha)$ is a unimodal curve. As a whole, the $f(\alpha)$ curves of different years are normal (Figure 5). The problems are not obvious in the graphs of local fractal dimension spectrums.

**Table 2 The estimation results of the capacity dimension, information dimension and correlation dimension based on 0-intercept regression and corresponding goodness of fit**

| Year | Capacity dimension | | Information dimension | | Correlation dimension | |
| --- | --- | --- | --- | --- | --- | --- |
| | $D_0$ | $R^2$ | $D_1$ | $R^2$ | $D_2$ | $R^2$ |
| 1984 | 1.7963 | 0.9991 | 1.7349 | 0.9988 | 1.7089 | 0.9964 |
| 1988 | 1.8213 | 0.9995 | 1.7749 | 0.9987 | 1.7570 | 0.9968 |
| 1989 | 1.8115 | 0.9988 | 1.7537 | 0.9992 | 1.7293 | 0.9972 |
| 1991 | 1.8060 | 0.9997 | 1.7555 | 0.9982 | 1.7356 | 0.9961 |
| 1992 | 1.7959 | 0.9995 | 1.7407 | 0.9988 | 1.7187 | 0.9968 |
| 1994 | 1.8184 | 0.9996 | 1.7781 | 0.9988 | 1.7619 | 0.9970 |
| 1995 | 1.8108 | 0.9993 | 1.7493 | 0.9985 | 1.7237 | 0.9957 |
| 1996 | 1.8159 | 0.9992 | 1.7548 | 0.9988 | 1.7282 | 0.9957 |
| 1998 | 1.8147 | 0.9995 | 1.7596 | 0.9986 | 1.7373 | 0.9963 |
| 1999 | 1.8228 | 0.9994 | 1.7669 | 0.9986 | 1.7445 | 0.9962 |
| 2001 | 1.8388 | 0.9993 | 1.7831 | 0.9990 | 1.7603 | 0.9972 |
| 2006 | 1.8297 | 0.9990 | 1.7601 | 0.9992 | 1.7296 | 0.9968 |
| 2009 | 1.8477 | 0.9992 | 1.7837 | 0.9992 | 1.7562 | 0.9974 |

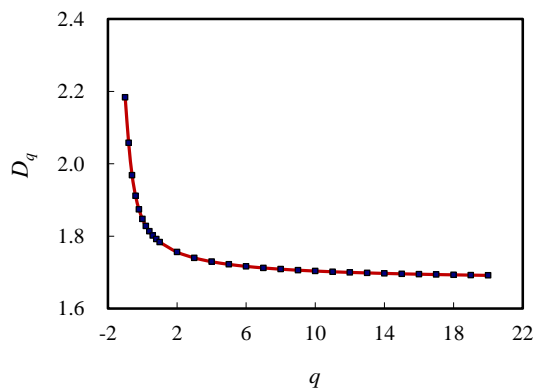
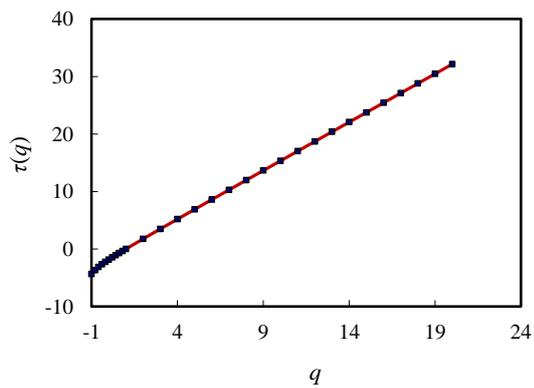

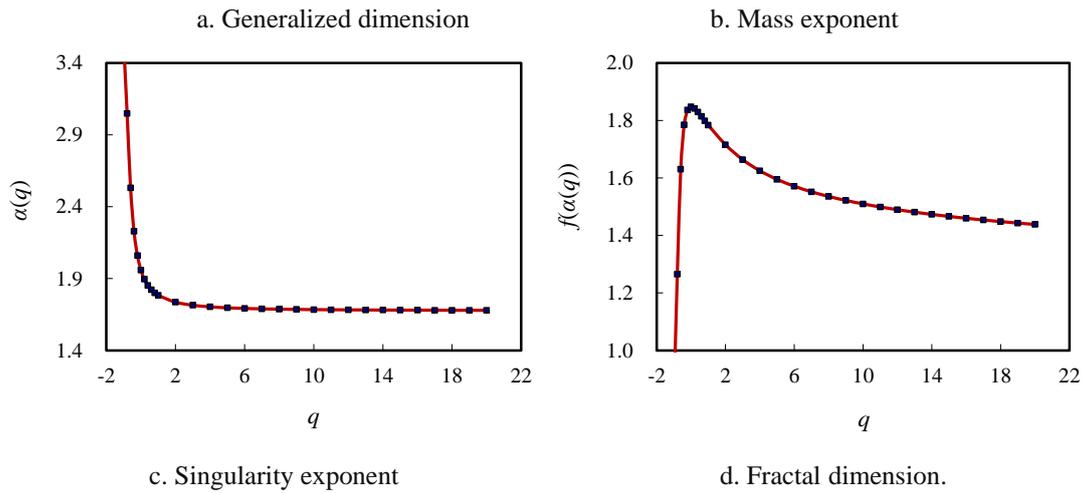

a. Generalized dimension  b. Mass exponent

c. Singularity exponent  d. Fractal dimension.

**Figure 4 The multifractal parameter spectrums of Beijing's urban form based on CCA and zero-intercept regression (2009)** [**Note:** The spectral curves are normal within the scales of moment order range from $q$=-0.5 to $q$=30]

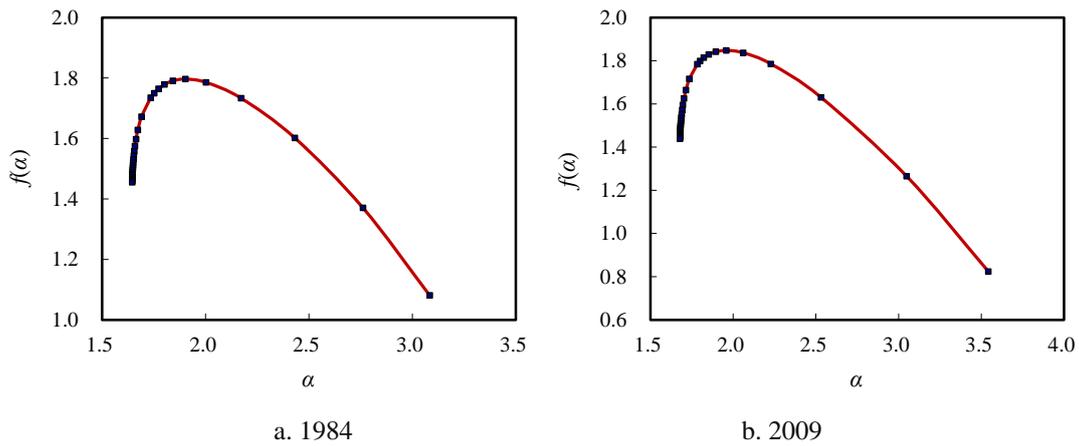

a. 1984  b. 2009

**Figure 5 The $f(\alpha)$ curves of Beijing's urban form based on CCA and zero-intercept regression (1984&2009)** [**Note:** The curves are normal within the scales of moment order range from $q$=-0.5 to $q$=30]

## 3.3 Calculations based on free intercept (0<$K$<2)

If we make a common regression analysis to estimate multifractal parameters, the results will be different to a degree from those shown above. In other words, if the intercept of the log-log linear regression equation is unconstrained, the fractal dimension values may be aberrant. The rationality of multifractal parameters can be preliminarily judged by the capacity dimension $D_0$, information dimension $D_1$, and correlation dimension $D_2$. According to equation (9), the reasonable numerical relationship between the three basic generalized correlation dimension values takes on an inequality: $D_0>D_1>D_2$. If the inequality is broken, the results are illogical and cannot be acceptable. If so, it suggests that there is probably something wrong in the process of calculation. Letting the intercepts $\ln K$, $I_0$, and $C_0$ free, we can make least square calculations using equations (14), (16), and (18) to estimate the capacity dimension, information dimension, and correlation dimension (Table 3). In this case, the goodness of fit becomes better, compared with the $R$ square of the first set of results (Table 2). However, the numerical relationships between $D_0$, $D_1$, and $D_2$ violate the monotonic decreasing principle of generalized correlation dimension. For example, for 1995 year, $D_0=1.7789<D_1=1.7793<D_2=1.7862$; for 2009 year, $D_1=1.8106<D_2=1.8133$. These are abnormal numerical relationships for multifractal measurements.

Table 3 The estimation results of the capacity dimension, information dimension and correlation dimension based on common regression and corresponding goodness of fit

| Year | Capacity dimension | | Information dimension | | Correlation dimension | |
|---|---|---|---|---|---|---|
| | $D_0$ | $R^2$ | $D_1$ | $R^2$ | $D_2$ | $R^2$ |
| 1984 | 1.7620 | 0.9997 | 1.7571 | 0.9990 | 1.7583 | 0.9975 |
| 1988 | 1.7951 | 0.9998 | 1.8010 | 0.9990 | 1.8061 | 0.9978 |
| 1989 | 1.7723 | 0.9995 | 1.7704 | 0.9994 | 1.7742 | 0.9981 |
| 1991 | 1.7865 | 0.9999 | 1.7885 | 0.9987 | 1.7921 | 0.9975 |
| 1992 | 1.7708 | 0.9998 | 1.7721 | 0.9992 | 1.7771 | 0.9983 |
| 1994 | 1.7965 | 0.9998 | 1.8078 | 0.9992 | 1.8155 | 0.9982 |
| 1995 | 1.7789 | 0.9997 | 1.7793 | 0.9989 | 1.7862 | 0.9974 |
| 1996 | 1.7830 | 0.9997 | 1.7803 | 0.9991 | 1.7861 | 0.9972 |
| 1998 | 1.7887 | 0.9998 | 1.7851 | 0.9989 | 1.7889 | 0.9974 |
| 1999 | 1.7942 | 0.9998 | 1.7976 | 0.9991 | 1.8038 | 0.9977 |
| 2001 | 1.8091 | 0.9997 | 1.8017 | 0.9992 | 1.8010 | 0.9979 |
| 2006 | 1.7938 | 0.9996 | 1.7880 | 0.9995 | 1.7930 | 0.9985 |
| 2009 | 1.8162 | 0.9997 | 1.8106 | 0.9995 | 1.8133 | 0.9988 |

The wrong multifractal dimension spectrums can be brought to light by graphs. Four multifractal parameter curves of 2009 year are displayed in Figure 6. The generalized dimension spectrum is not normal. If $q>0$, the $D_q$ values fail to decrease monotonically; If $q<-1$, the $D_q$ values fail to converge (Figure 6a). The mass exponent cannot reflect problem clearly. The $\tau_q$ values seem to linear change (Figure 6b). The case of the singularity exponent is similar to the generalized dimension. If $q>0$, the $\alpha(q)$ values do not decrease monotonically; If $q<-1$, the $\alpha(q)$ values fail to approach a limit (Figure 6c). The curve of the local fractal dimension is apparently improper. This is supposed to be a curve with a single peak. The maximum value of the $f(\alpha)$ should equal the capacity dimension $D_0$. However, in these calculations, many $f(\alpha)$ values are greater than $D_0$ when $q>0$ (Figure 6d). The problems can be clearly reflected by the relationship between $\alpha(q)$ and $f(\alpha)$ curve, which is expected to be a unimodal curve. However, the curve shape is weird and it is hard to interpret this trend logically (Figure 7).

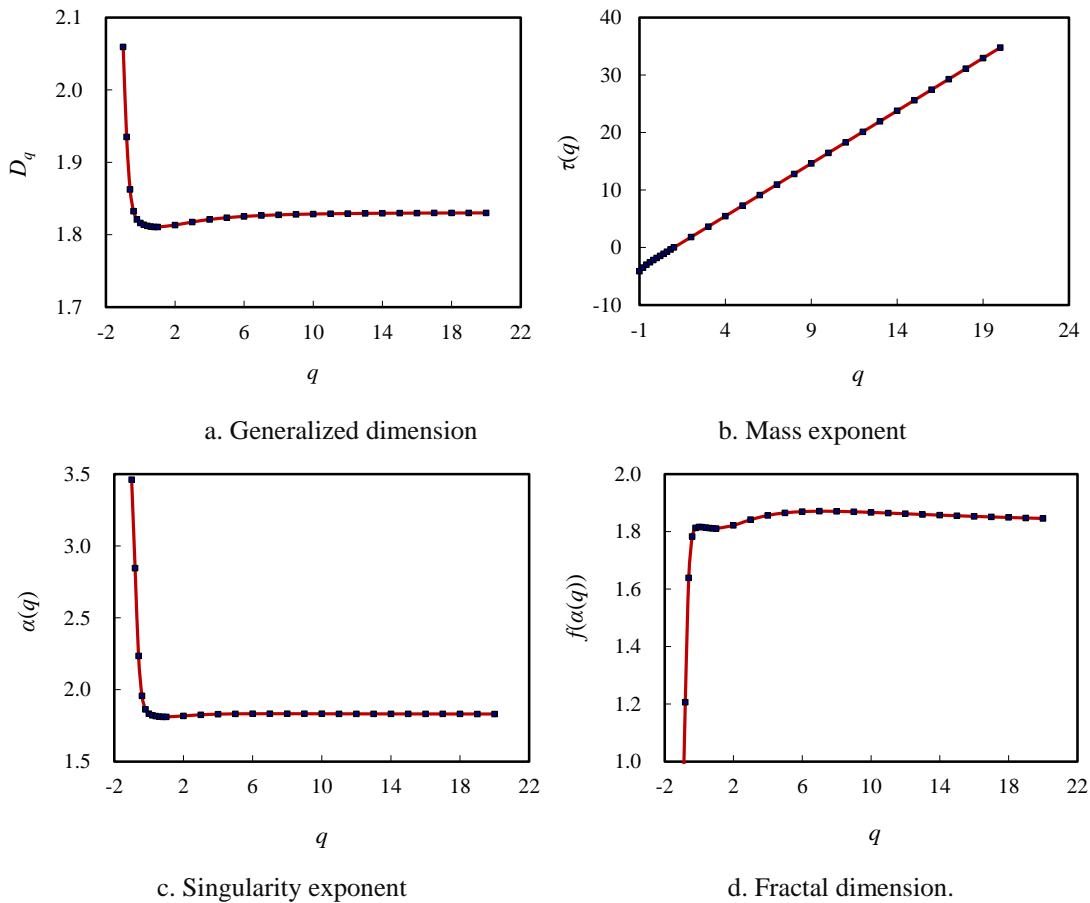

a. Generalized dimension  
b. Mass exponent  
c. Singularity exponent  
d. Fractal dimension.

**Figure 6 The abnormal multifractal parameter spectrums of Beijing's urban form based on CCA and common regression (2009)** [**Note:** The spectral curves are abnormal within the scales of moment order range from $q=-0.5$ to $q=30$]

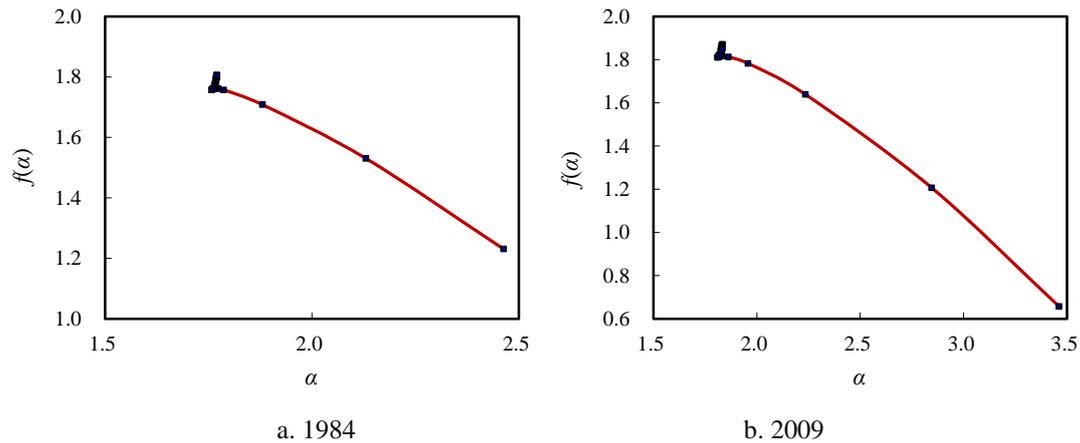

a. 1984                                       b. 2009

**Figure 7 The abnormal *f(α)* curves of Beijing's urban form based on CCA and common regression (1984&2009) [Note:** The curves are abnormal within the scales of moment order range from *q*=-0.5 to *q*=30**]**

# 4 Discussion

The log-log linear regression on the base of OLS is a basic approach to calculating multifractal parameters spectrums, but different ways of application lead to different results. Comparing and contrasting the results based on the zero-intercept regression (Table 2) and those based on the conventional regression (Table 3), we can find the more advisable approach to calculating multifractal parameters of urban form. If the intercept of the log-log linear equation is fixed to zero, the multifractal parameter values will be normal within certain range of scales and form proper curves (Figures 4 and 5). In contrast, if the intercept of the regression model is free, the multifractal parameter spectrums may be abnormal and take on improper curves (Figures 6 and 7). Cities are not real fractals, but they can be treated as random pre-fractal objects. If an object presents fractal properties within certain scaling range, it can be regarded as a pre-fractal system (Mandelbrot, 1982). As indicated above, the multifractal parameter values of cities, including the global parameter $\tau_q$ and $D_q$ and the local parameters $\alpha(q)$ and $f(\alpha)$ are right only within certain range of moment order (*q*). A comparison can be drawn between the multifractal measurement effect based on the zero-intercept regression and that based on the conventional regression (Table 4).

**Table 4 Comparison between the multifractal measurement effect based on the zero-intercept regression and that based on the conventional regression**

| Level | Parameter | Zero-intercept regression (constrained intercept) | Conventional regression (unconstrained intercept) |
|---|---|---|---|
| Global level | Generalized correlation dimension $D_q$ | Monotonic decreasing quantity (normal) | Non-monotonic decreasing quantity (abnormal) |
| | Mass exponent $\tau_q$ | Monotonic increasing quantity (normal) | Monotonic increasing quantity (seeming normal) |
| Local level | Singularity exponent $\alpha(q)$ | Monotonic decreasing quantity (normal) | Non-monotonic decreasing quantity (abnormal) |
| | Local fractal dimension $f(\alpha)$ | Single-peak curve: first increasing and then decreasing quantity (normal) | Convex curve: fluctuant quantity (abnormal) |

**Note**: For the zero-intercept regression, the constant term is fixed to zero ($\ln K=0$), and thus the proportionality coefficient is $K=1$. The goodness of fit is measured by squared cosine. For the conventional regression, the constant term is free, and thus the proportionality coefficient is an arbitrary number. The goodness of fit is defined as squared correlation coefficient.

The calculation effect of multifractal measurements depends not only on the methods of log-log linear regression but also on the fractal property or fractal development. For the regular multifractals, or the well-developed growing multifractals, no matter what methods are employed, the fractal parameter values are proper within certain range of moment order (Chen, 2014). If the multifractal structure is less developed, no matter what methods are used, the parameter values are improper. However, for the developing multifractal structure, the multifractal measurement effect depends on approaches. If the method is proper, the results will be normal, or else the results will be abnormal. Based on the ideas above-mentioned, multifractal cities can be divided into three categories (Table 5). And thus, multifractal theory can be employed to judge the development stages of cities. If the reasonable multifractal parameter spectrums be independent of calculation methods, the multifractal structure of a city is well developed; if the reasonable multifractal dimension spectrums depend on calculation approaches, the multifractal structure of the city is developing or less developed; if any methods result in absurd multifractal spectral curves, the multifractal structure of the city has not yet developed at all.

**Table 5 Classification of fractal cities based on multifractal measurement approaches and effect**

| Type | Fractal structure | Multifractal measurement effect |
|---|---|---|
| Multiscaling ideal fractal cities | Well developed fractal structure | No matter what approaches are used, the parameter values are normal |
| Multiscaling pre-fractal cities | Developing fractal structure | Multifractal measurement effect depends on the approaches used |
| Potential fractal cities | Less developed fractal structure | No matter what approaches are used, the parameter values are improper |

This is a study on methodology of multifractal dimension estimation rather than a case study of fractal cities, thus the empirical analysis is not the aim of this work. Despite that, we can reach significant conclusion on the urban structure of Beijing city. As we know, the order of moment $q>1$ reflects the region of measurement denseness, while the moment order $q<0$ reflect the region of measurement sparseness (Wang and Li, 1996). If the $q$ value approaches positive infinity ($q \to +\infty$), the multifractal measure tends to the higher-density region of a city, corresponding to the higher probability of urban growth; on the contrary, if the $q$ value approaches negative infinity ($q \to -\infty$), the multifractal measure tends to the lower-density region of the city, indicating the process of lower growth probability (Chen, 2014). As far as Beijing is concerned, when $q \to +\infty$, the multifractal dimension spectrums are normal, but when $q \to -\infty$, the multifractal parameter spectrums are abnormal (Figures 4 and 5). This suggests that the multifractal patterns of the densely populated areas of Beijing has developed, but the sparsely populated areas bear no multifractal pattern (Chen and Wang, 2013). Fractals imply a kind of optimized structure of nature. A fractal object can occupied its space in the most efficient way (Chen, 2016). Using the ideas from fractals to design cities will help human being make the most of geographical space. In this sense, the theory and method of fractal city planning should be developed in order to improve the internal spatial structure and external service function of Beijing.

The zero-intercept log-log linear regression is not the only approach to yield proper multifractal parameter spectrums. Another feasible approach is to extend scaling range (Chen, 2014). On a double logarithmic plot for fractal dimension estimation, the scattered points displaying the numerical relationship between the linear size of boxes and the number of nonempty boxes can be divided into three scale ranges. In fact, the first scale range corresponds to the Euclidean dimension of embedding space ($d=2$), and the third scale range corresponds to the topological dimension of a fractal set ($d_T=0$), only the second scale range reflects the fractal structure with a fractal dimension ($0<D<2$). In other words, among the three scale ranges, the middle part (the second scale range) represents scale-free range and is always termed scaling range (Chen, 2011; Mandelbrot, 1982; Williams, 1997). The slope of the line segment based on the scaling range gives a fractal parameter value. In monofractal analysis, the fractal dimension estimation is always based on the scaling range, namely, the second scale range. For multifractal measurement, in order to obtain proper parameter values, we can combine the first scale range and the second scale range to estimate multifractal parameters. This kind of treatment is in effect similar to fixing the intercept of a log-log linear regression model to zero, and the multifractal dimension spectrums are often normal (Chen, 2014). In short, for the developing fractal cities, the proportionality coefficient of the power function indicative of fractal structure should be constrained to the unit ($K$=1) by some method so that the fractal parameters take on proper values.

The deficiency of this study lies in three aspects. First, only the OLS method are examined, other algorithms such as maximum likelihood estimation (MLE) and major axis (MA) method are not investigated. A conjecture is that the conclusions of this work can be generalized to other algorithms. After all, an algorithm is just a step-by-step problem-solving procedure, and different algorithms have no essential difference. Second, the definition of study area impacts on multifractal parameter estimation. Changing size of a study area, the effect of multifractal measurement may be different. Third, the searching radius influences the effect of multifractal parameter estimation. In practice, the length of searching radius should be associated with resolution of remote sensing images. Because of the limited space of this paper, the related problems remain to be solved in the future.

## 5 Conclusions

This paper is devoted to exploring the methods of multifractal parameter estimation using OLS regression. Two ways of log-log linear regression analysis are compared with each other. One is to fix the constant term to zero, namely, zero-intercept regression; the other is not to fix the constant term to zero, that is, conventional regression. Based on fractal theory, the remote sensing data of Beijing city from 1984 to 2009 is employed to make empirical analysis. Comparing the similarities and differences between the two sets of results, we can draw a number of conclusions.

The main conclusions of this study are as follows. **First, only proper approach to evaluating multifractal parameters of cities can result in normal multifractal dimension spectrums.** Where linear regression analysis is concerned, the more advisable approach is to constrain the constant term to zero. If the intercept is fixed to zero in the double logarithmic linear regression, we can obtain more reasonable multifractal parameter values. In contrast, if the intercept is free in the log-log linear regression, we may get abnormal multifractal dimension values. Thus, the fixed-intercept method is preferable when calculating multifractal spectrums of urban patterns. The simplest method of judging the rationality of the multifractal dimension sets is to compare the values of capacity dimension, information dimension, and correlation dimension. If the capacity dimension is greater than the information dimension, and the information dimension is greater than the correlation dimension, the results are logical, otherwise, there may be something wrong in the process or results of calculation and analysis. **Second, in urban study, the two different approaches can be combined to make an empirical analysis of multifractal structure.** The problems of multifractal parameter spectrums may be caused by different factors. For a regular and typical multifractal object, the parameter estimation results are independent of computational approaches. That is, the methods of zero-intercept regression and conventional regression make no significant difference. However, if the multifractal structure is less developed, the results will depend on the methods. In this case, the zero-intercept regression approach may yield more reasonable results than the common regression. Consequently, if a problem of a multifractal dimension spectrum of a city is independent of computational methods (say, when $q<-1$, the multifractal parameter values of Beijing urban form are abnormal, and these do not depend on the calculation methods), the problem is really related to the urban structure and the city should be improved by city planning and urban design. **Third, an inference is that the proportionality coefficient of the power laws indicative of multifractals should be fixed to unit in order to obtain proper multifractal parameter spectrums.** To make a linear regression analysis based on the least squares calculation, a power function should be transformed into a log-log linear relation, in which the proportionality coefficient is converted into a constant term (intercept). However, if we employ the nonlinear least-squares regression technique to estimate fractal parameters, it will be unnecessary to transform the power function into a linear function, and there will be no intercept in the regression equation. In this case, the proportionality coefficient of the power function can be constrained by the unit ($K$=1).

**Acknowledgements**

This research was sponsored by the National Natural Science Foundations of China (Grant No. 41590843 & 41671167). The supports are gratefully acknowledged.

# Appendices

### Appendix 1. The limit of 0 power of 0 approaches to 1

To understand the derivation of the fractal model with unitary proportionality coefficient, we should discuss the 0 power of 0, namely, $0^0$. Let

$$y = x^x, \tag{A1}$$

where $x$ approaches to 0. Then we have

$$\ln y = x \ln x = \frac{\ln x}{\frac{1}{x}}. \tag{A2}$$

By l'Hospital's principle, we have

$$\ln y \to \lim_{x \to 0} \frac{\frac{1}{x}}{-\frac{1}{x^2}} = -x \to 0, \tag{A3}$$

which indicates

$$y \to 1. \tag{A4}$$

This derivation is helpful for understanding equation (4) based on equation (3).

**Appendix 2. How to use Legendre transform to estimation multifractal parameters**

Using Legendre transform, we must turn the differential equation into difference equation. In other words, in order to estimate $\alpha(q)$ and $f(\alpha)$, equation (7) should be discretized as below

$$\alpha(q) = \frac{\Delta \tau(q)}{\Delta q} = D_q + (q-1)\frac{\Delta D_q}{\Delta q}, \tag{B1}$$

where $\Delta$ represents difference operator. Accordingly, equation (8) can be expressed as

$$f(\alpha) = q\frac{\Delta \tau(q)}{\Delta q} - \tau(q) = q\frac{\Delta \tau(q)}{\Delta q} - (q-1)D_q . \tag{B2}$$

Using equations (B1) and (B2), we can estimate the local parameters of multifractals. The smaller $\Delta q$ is, the exacter $\alpha(q)$ and $f(\alpha)$ will be.